 \documentstyle[prl,aps,multicol,epsf,graphicx]{revtex}

\begin{document}



\title {Oscillatory Instabilities of Standing Waves in One-Dimensional 
Nonlinear Lattices}

\author{Anna Maria Morgante, Magnus Johansson, Georgios Kopidakis, 
 and Serge Aubry}

\address{Laboratoire L\'eon Brillouin (CEA-CNRS),
   CEA Saclay,
   F-91191 Gif-sur-Yvette Cedex,
   France}
\date{April 18, 2000}
\maketitle

\begin{abstract}
In one-dimensional anharmonic lattices, we construct 
 nonlinear 
standing waves (SWs) reducing to harmonic SWs at small amplitude. 
For SWs with spatial periodicity incommensurate with the lattice period, a
transition by breaking of analyticity versus wave amplitude is observed. 
As a consequence of the discreteness, oscillatory linear
instabilities, persisting for arbitrarily small 
amplitude in infinite lattices, appear for all wave numbers $Q \neq 0, \pi$. 
Incommensurate analytic 
SWs with  $|Q|>\pi/2$ may however appear as 'quasi-stable', as their 
instability 
growth rate is of higher order. 
%
\end{abstract}
\draft 
\begin{multicols}{2}
A wellknown and rather spectacular phenomenon occuring in many nonlinear 
 media (e.g. fluids or optical wave\-guides) is the modulational 
(Benjamin-Feir)
instability (MI), 
by which a  travelling plane wave breaks up into a train of solitary waves 
(see e.g. \cite{IR}). 
It is also wellknown that wave propagation in many continuous nonlinear media 
is well described by Nonlinear Schr\"odinger type equations, where the 
solitary  wave trains are described by spatially periodic and stable 
 {\em standing wave} 
(SW) solutions, the so-called cnoidal envelope waves \cite{AEK}. 

An analogous instability of propagating waves occurs also in {\em discrete} 
systems (e.g. 
anharmonic lattices), where in the case of soft (hard) anharmonicity, waves 
with small (large) wave numbers are unstable \cite{Peyrard,CA} and typically 
break up into arrays of intrinsically localized modes or {\em discrete 
breathers} (see e.g. \cite{Aub97,Flach}). However, concerning the possible 
existence and stability of spatially periodic SWs in anharmonic lattices, 
much less has 
been 
known up to now. Although stable SW solutions of the form 
$A \cos(Qn)\cos(\omega t)$ generally  exist in harmonic lattices as linear 
superpositions of  counterpropagating  waves with equal
amplitudes $A$ and frequencies $\omega$ but opposite wave numbers $\pm Q$, 
there is {\it a priori} no guarantee that these SWs will remain stable in the 
presence of anharmonicity. 

Here, we consider the general 
class of one-dimensional (1D) anharmonic lattices described by a discrete 
nonlinear Klein-Gordon (KG) or Schr\"odinger (DNLS) equation. We first 
propose  a method to generate (numerically) exact time-periodic 
SW solutions to these equations. These SWs, whose 
spatial periodicity can be either commensurate
($Q/2\pi$ rational) or
incommensurate ($Q/2\pi$ irrational) with the lattice period, are found 
as {\em 'multibreather'} solutions \cite{Aub97} from the so called 
{\em 'anticontinuous'} limit of zero inter-site coupling, and they reduce to 
the 
harmonic SWs for small amplitudes. Then, we investigate the 
stability properties of these SWs, and find as a most striking
result that for  infinite lattices, SWs with 
{\em small but nonzero} amplitude are
{\it unstable through
an oscillatory instability for all $0<Q<\pi$ } \cite{Marin98}. Thus, the SWs 
are unstable also for wave numbers where MI for 
propagating waves does not occur. 
It is important to note that the additional
SW instabilities appear as a direct consequence of the lattice
discreteness, and have not been found within continuum approximations 
\cite{KHS}. Moreover, the dynamics resulting from the SW instabilities is 
fundamentally different from that of the MI, as the latter generally is 
non-oscillatory (characterized by purely imaginary  eigenfrequencies  
of the linearized 
equations (\ref{eigeneqharm2}) below) while the former is oscillatory 
(corresponding 
to complex eigenfrequencies with nonzero real part) \cite{dark}.  In this Letter, our 
main results will be stated and motivated
briefly; full technical details will be given elsewhere \cite{Anna}.

We thus consider a 1D chain of classical anharmonic oscillators with 
a general 
on-site potential $V(u)$ and harmonic inter-site coupling 
$C_{K}>0$, yielding 
a discrete nonlinear KG equation for the 
particle displacements $u_n$, 
\begin{equation}
\label{DKG}
\ddot{u}_n + V'(u_n)  - C_K(u_{n+1} + u_{n-1} -2 u_n) = 0 .
\end{equation}
Here, we shall be mainly concerned with  the small-amplitude dynamics of 
KG chains with small 
inter-site coupling $C_K$, which can be well described by a DNLS approximation 
\cite{Peyrard,KHS,Anna} (its range of validity is 
investigated in more details in \cite{Anna}). 
For small oscillations, the on-site potential $V(u)$ can be expanded  
as
$V(u)$ = $u^2/2$ $+$ $\alpha u^3/3$ $+$ $\beta u^4/4$ $+ ...$, where
the linear oscillator frequency is set to 1. Linearizing
Eq.\ (\ref{DKG}) yields propagating or SW solutions with  wave number $Q$ 
and frequency $\omega_0(Q)$ given by the 
dispersion relation  $\omega_0^{2}(Q) = 1+4C_K \sin^2 {Q/2}$. For 
the nonlinear Eq.\ (\ref{DKG}), we search for small-amplitude time-periodic 
solutions as $u_n(t) = \sum_p
a^{(p)}_n e^{i p \omega_b t }$, for which the dynamics will be 
almost harmonic with frequency $\omega_{b}$ 
close to $\omega_{0}(Q)$ for some $Q$. Then, 
$\omega_b^2-1 \equiv  2 \delta $ will be of order $C_K$, and 
allowing for a slow time dependence of 
the Fourier coefficients $a_n^{(p)}$ yields \cite{Anna}, at order $C_K$,  
a DNLS equation 
for the dominating 
coefficient $a_n^{(1)}$ describing the leading-order nonlinear effects. 
Defining 
$\lambda$ $=$ $-5\alpha^2/3$ + $3\beta/2$ (so that for soft [hard] potentials,
$\lambda < 0 $ [$\lambda > 0$])
 and $\psi_n=\sqrt{|\lambda|} a_n^{(1)}$, 
the DNLS equation reads
\begin{equation}
 i \dot{\psi}_{n}=\delta \psi_{n}-\sigma |\psi_{n}|^{2}
\psi_{n}+C(\psi_{n+1}+\psi_{n-1}-2 \psi_{n}) , 
 	\label{DNLS}
 \end{equation}
with  $\sigma={\rm sign}(\lambda)$ and  $C=C_K/2$. 
We will assume $\sigma=-1$ without loss of generality,
since changing $\psi_n \rightarrow (-1)^n \psi_n^{\ast}$, $\delta \rightarrow
 4 C-\delta $ is equivalent to changing the sign of $\sigma$.

For time-periodic solutions  all $a^{(p)}_n $, and 
therefore $\psi_n$, will be time-independent and can be chosen real. 
Then, vanishing the left-hand side 
of Eq.~(\ref{DNLS}) defines a nonlinear symplectic map ${\mathcal S}:
(\psi_{n},\psi_{n-1}) \rightarrow (\psi_{n+1},\psi_{n})$ in the 2D real 
plane, which, with the change of scale 
$\psi_{n} \rightarrow \sqrt{C}
 \psi_{n}$, depends only on the parameter
$\delta^{\prime} \equiv \delta/C$. As is wellknown \cite{Wan},
this map exhibits a rich variety of orbits,
including elliptic and hyperbolic fixpoints and periodic cycles, KAM tori,
Aubry-Mather Cantor sets (Cantori) \cite{Aub78} and chaotic orbits. 
Searching for  nonlinear SWs in continuation of 
the linear SWs $\psi_n = A \cos(Q n + \phi)$, with 
$\delta = 4 C \sin^2 {Q/2} \equiv \delta_0(Q)$,  only the periodic and 
quasiperiodic 
orbits  which can be continued to zero amplitude by varying 
$\delta^\prime$ are of interest. For small amplitudes, these orbits  
must be located close to the elliptic fixpoint $F_{0}=(0,0)$, 
and $Q$ must be close to its linear rotation number $\theta(\delta^\prime)$ 
given by $\delta =  \delta_0(\theta)$.  As $\sigma<0$, 
the rotation number for  orbits rotating around $F_{0}$ 
increases with increasing  radius, and since the map 
${\mathcal S}$ as well as its orbits at fixed rotation number depends 
continuously on $\delta^{\prime}$, {\em SWs with wave number $Q$ exist 
only for 
$\delta \leq \delta_{0}(Q)$}. 

When $Q= 2 \pi r/s$ ($r$ and $s$ irreducible 
integers), there are {\em two families of commensurate 
SWs}, each of them 
represented by one of the two $s$-periodic cycles (hereafter  called
{\em 'h-cycle'} resp.\ {\em 'e-cycle'}) which bifurcate in pair from $F_{0}$ 
at 
$\delta=\delta_{0}(Q)$  
\cite{Wan,Aub78,Aub92}. The h-cycle is hyperbolic 
for all $\delta^\prime$, while the e-cycle is elliptic for 
$\delta$ close to $\delta_{0}(Q)$ and hyperbolic with reflection below a 
 critical value $\delta = \delta_{c}(Q)$. 
  When $Q/2\pi$ is a generic irrational number, the trajectories representing 
{\em incommensurate} SWs emerge from $F_{0}$ as the KAM torus with rotation 
number $Q$.  
At the critical value $\delta = \delta_{c}(Q)$, the KAM torus breaks up in 
a {\em transition by breaking of analyticity} \cite {Aub78} and bifurcates 
into a Cantorus and its associated  'midgap' 
trajectory \cite{midgap}. Thus, for $\delta < \delta_{c}(Q)$ 
the Cantorus and the midgap trajectory define {\em two families of 
non-analytic 
incommensurate SWs}, which 
merge into the {\em unique analytic SW} defined by the KAM torus for 
$\delta \geq \delta_{c}(Q)$. 

To give a global representation of these  SWs,  
we consider the large-amplitude limit 
$\delta \rightarrow - \infty$, 
or, 
equivalently, the {\em anticontinuous} limit $C=0$ for fixed 
$\delta < 0$. 
From this limit, where stationary solutions to Eq. (\ref{DNLS}) can only take 
the values 
$\psi_n=\pm \sqrt{-\delta}$ and $\psi_n = 0$, 
we can describe the SWs as {\it multibreathers} 
\cite{Aub92,Aub97} 
identified by a {\em coding sequence} 
$\sigma^{\prime}_n$ defined as 
$\psi_n=\sqrt{-\delta} \sigma^{\prime}_n$. 
By continuing the  SW map orbits to large negative $\delta^{\prime}$,
we find their coding sequences as 
$\sigma^{\prime}_{n}=\chi_0(Q n+\phi)$, where 
$\chi_{0}(x)$ is the $2\pi$-periodic odd function defined for 
$x \in [0,\pi]$ as
\begin{equation}
\chi_0(x) = \left \{ \begin{array}{l} 1  \quad \mbox{for} \quad (\pi-Q)/2
\leq x \leq (\pi+Q)/2
 \\ 0 \quad \mbox{elsewhere} \end{array} \right.  .
\label{hull}
\end{equation}
For all phases  $\phi \neq \phi_m \equiv  \pm(\pi-Q)/2 -mQ$ 
($m$ integer), this describes SWs 
 corresponding  either to the h-cycles (for rational  $Q/2\pi$)
or to the Cantori (for irrational  $Q/2\pi$). We will call these SWs 
{\em 'type H'}, and they have the property (of importance for the 
stability analysis below) that their coding sequences do not contain any 
consecutive $+1$ or $-1$. 
For the particular phases $\phi = \phi_m$,
$x=Qm+\phi$ is at a discontinuity point of $\chi_{0}(x)$, so that 
$\sigma^{\prime}_n$ has two consecutive $+1$ (or $-1$). These SWs, 
called {\em type E}, correspond to e-cycles or to midgap 
trajectories. 

\begin{figure}[htbp]
\noindent
\centerline{\includegraphics[width=3.5cm,angle=0]{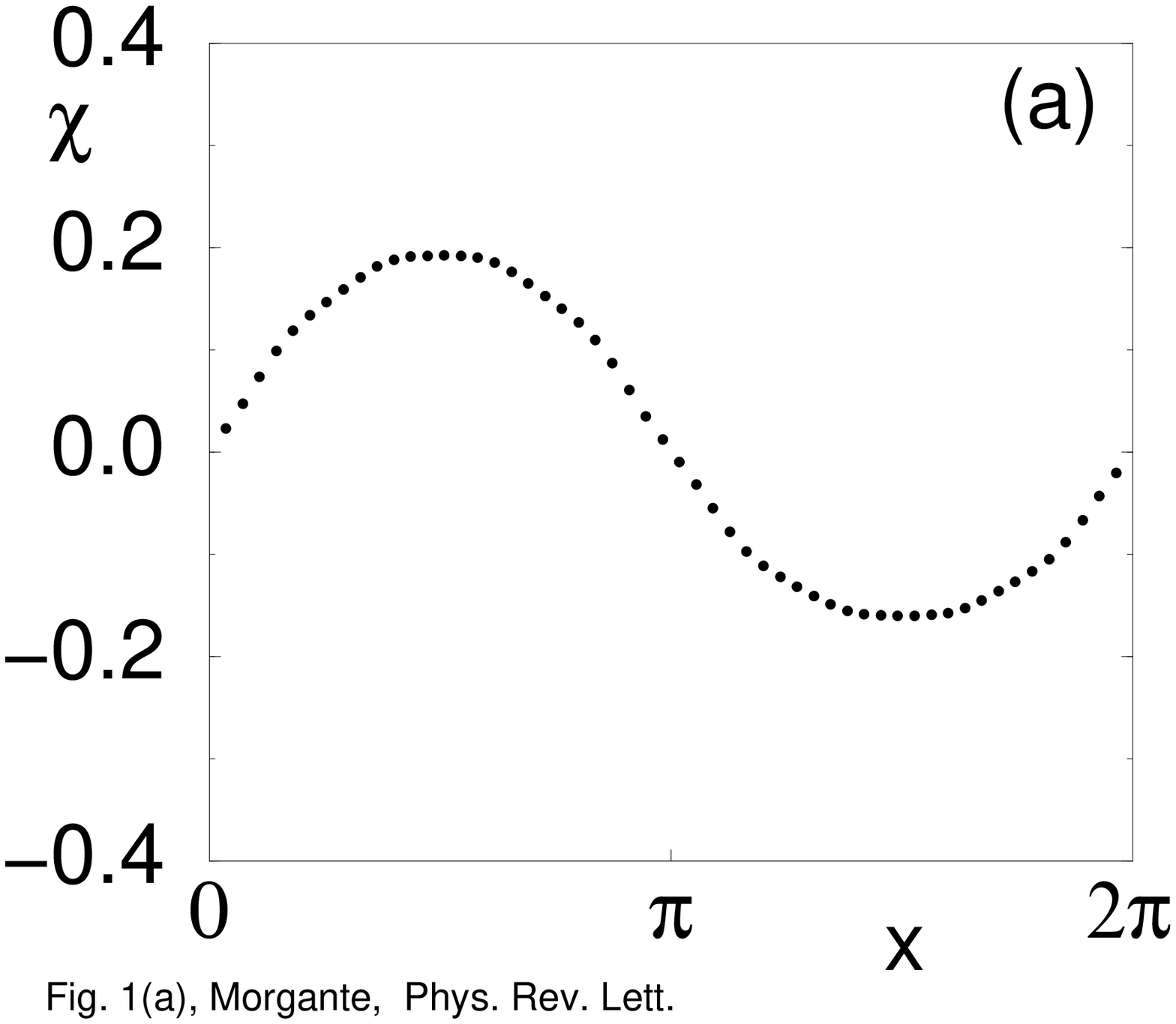}
\includegraphics[width=3.5cm,angle=0]{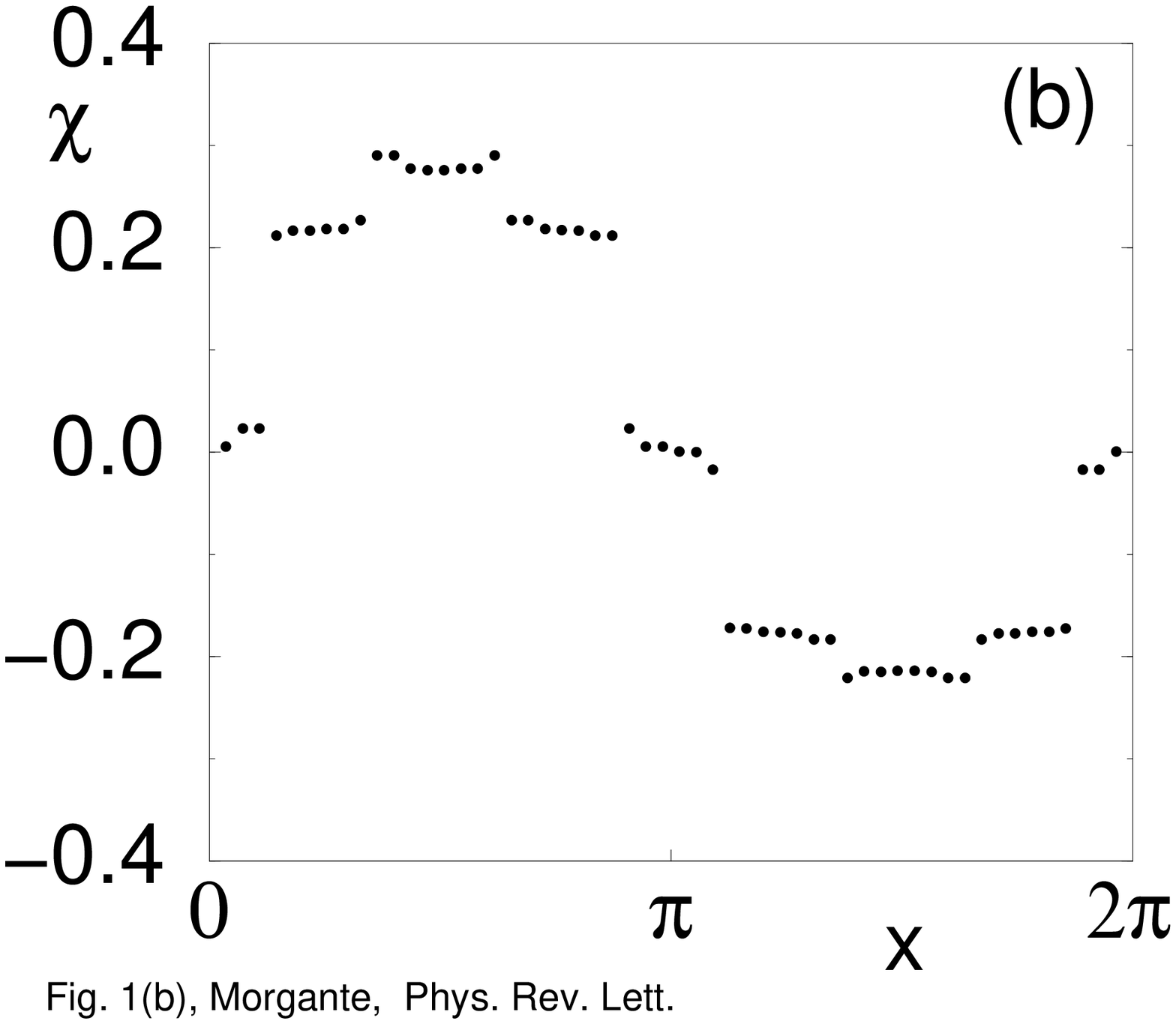}}
\caption{Hull functions $\chi(x)$ for SWs in a KG chain (\ref{DKG}) with 
$V(u)=\frac{1}{2}\left(e^{-u}-1\right)^2$, $C_K=0.05$, 
$Q/2\pi = 21/55$ $\simeq$ $ (3-\sqrt{5})/2$ $\equiv$ $ \sigma_G$ and (a) 
$\omega_{b}=1.072$ resp.\ (b) $\omega_{b}=1.060$.  
The breaking of analyticity occurs at  $\omega_{b} \simeq 1.068$ 
for $Q/2\pi = \sigma_G$.
\label{transition}}
\end{figure}

The representation (\ref{hull}) is useful also for numerically 
calculating directly the exact SWs of the original KG equation 
(\ref{DKG}), using standard methods 
(e.g.\ Newton schemes) for continuation of multibreathers from the 
anticontinuous limit. 
Continuing these solutions  versus $\omega_{b}$ to zero amplitude, we 
typically find  (Fig.~\ref{transition}) a transition 
by breaking of analyticity of the hull function $\chi(x)$, defined as 
$u_{n}(0) = \chi(Qn+\phi)$, for incommensurate SWs in the KG chains as well as 
in DNLS chains.

We now turn to the investigation of  the dynamical stability properties 
of these SWs. As before, we consider small-amplitude solutions to the KG 
equation (\ref{DKG}) with small $C_K$, so that the DNLS approximation 
(\ref{DNLS}) is well justified.
With the substitution  $\psi_n \rightarrow \psi_{n}+\epsilon_{n}(t)$, where 
$\psi_n$ is real and time-independent and  $\epsilon_{n}$ is small, 
the linearization of Eq.~(\ref{DNLS}) yields a standard Hill equation 
for $\epsilon_{n}$. Its eigenmodes
can be found
by setting  $\epsilon_{n}(t)=\frac{1}{2}(u_{n}+v_n) e^{i \omega_l t} +
\frac{1}{2} (u_{n}^{*}-v_n^*) e^{-i \omega_l t}$, where the eigenfrequencies
$\omega_l$
are given by the eigenvalue problem:
\begin{eqnarray}
\nonumber
{\cal L}_0 v_n \equiv  (2C-\delta - \psi_{n}^{2}) v_{n}-C(v_{n+1}+v_{n-1})
&=&  \omega_l u_{n} \\
{\cal L}_1 u_n \equiv (2C-\delta -3 \psi_{n}^{2}) u_{n}-C(u_{n+1}+u_{n-1})&=&
 \omega_l v_{n} .
\label{eigeneqharm2}
\end{eqnarray}
The SW $\psi_n$ is linearly stable if and only if all eigenfrequencies
$\omega_l$ are real. 

\begin{figure}[htbp]
\noindent
\includegraphics[height=7cm,angle=270]{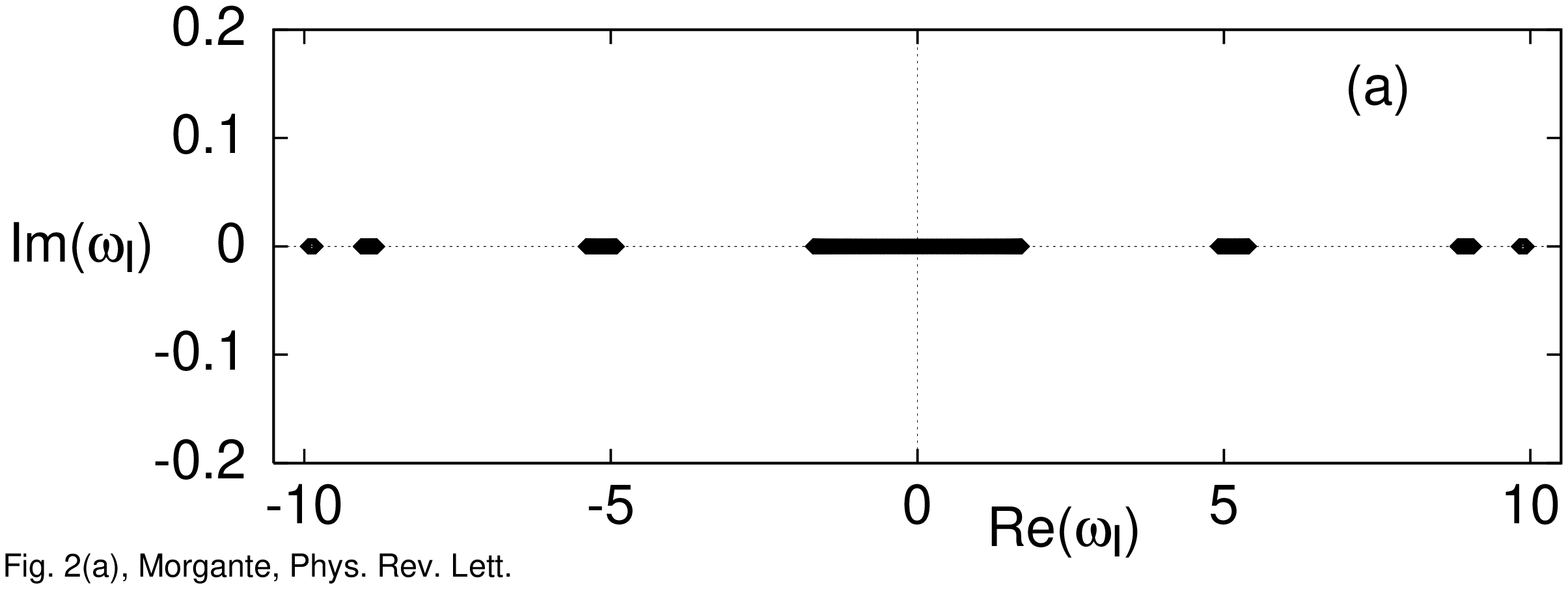}
\\
\includegraphics[height=7cm,angle=270]{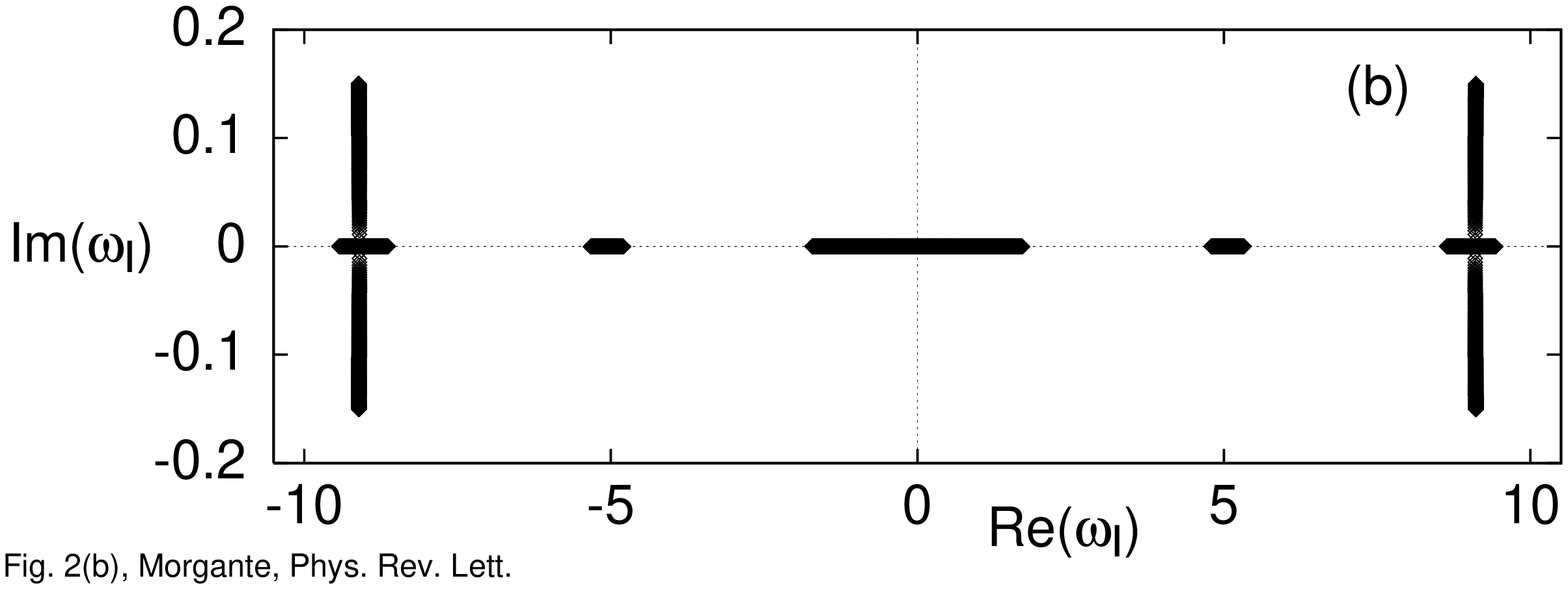}
\caption{Eigenvalues of Eq.\ (\ref{eigeneqharm2}) for a type H SW with 
$Q=3 \pi / 4 $ and (a) $\delta^\prime=-8.0$ resp. (b) 
$\delta^\prime=-7.5$ ($C=1$). }
\label{collision}
\end{figure}

At the anticontinuous limit ($\delta <0$, $C=0$), 
the eigenvalues of (\ref{eigeneqharm2}) will be located either at 
$\omega_l=0$ (corresponding to sites with codes 
 $|\sigma_n^\prime|=1$) or at $\omega_l=\pm \delta$ (corresponding to 
$\sigma_n^\prime=0$).  
For commensurate SWs ($Q=2 \pi r/s$)
the lattice periodicity of $\psi_{n}^{2}$ in (\ref{eigeneqharm2}) implies
that the degenerate eigenvalues will form bands as $C$ is increased from zero
(see Fig. \ref{collision}). There will be $s-2r$  pairs 
of bands originating from $\omega_{l}= \pm\delta$, and  $r$  pairs of bands 
originating from $\omega_{l}= 0$ (which always remains in the spectrum). 
For incommensurate SWs, the band structure 
will be Cantor-like. In all cases, bands originating from 
$\pm \delta$ can be shown to move initially only along the 
real axis when $C$ is increased from zero, while the behaviour of the 
bands  originating from $0$ will 
be different for E and H type SWs. As the coding sequences for 
E type SWs 
contain pairs of consecutive $+1$ (or $-1$) and the anharmonicity is soft, 
it follows from a general result \cite {Aub97} that some 
eigen\-values always move out on the imaginary axis as soon as 
$C\neq 0$, 
making these SWs {\it dynamically unstable}  for $\delta<0$ and small $C$. 
Moreover, we find \cite{Anna} that {\em this instability persists for
all $\delta < \delta_{0}(Q)$}, so that SWs of type E are unstable for any 
nonvanishing amplitude. By contrary, for H type SWs 
the theory of effective action \cite{Aub97} can be used to show that 
all eigenvalues  initially will move along the real axis, and 
thus {\em the stability of these SWs is preserved for $C$ not too large}
(Fig.\ \ref{collision} (a)).

Increasing $\delta$ (for fixed $C$) the bands 
broaden, 
and the main gap separating the two classes of bands originating from $0$ 
resp.\ $\pm\delta$ shrinks. 
At some value $\delta$ $=$ $\delta_{K}(Q)$ it vanishes, and, as shown in 
Fig.\ \ref{collision}(b), eigenvalues from the 
two classes of bands will collide and move out in the 
complex plane (since they have opposite 
{\em Krein signature} \cite{Aub97}). Thus, {\em oscillatory instabilities 
appear 
 for the SW}. These instabilities have also been confirmed by 
direct numerical Floquet analysis of the KG Eq.\ (\ref{DKG}) for small 
$C_K$ \cite{Anna}.
\vspace{-0.5cm}
For $\delta > \delta_{K}(Q)$ bands continue to overlap, generating 
new instabilities. 
In many cases, the SW remains unstable
until it vanishes at $\delta=\delta_0(Q)$, although  for some $Q$ 
with small $s$  the 
stability is temporarily regained for some interval of $\delta$ where no 
bands overlap. But for {\em all} commensurate SWs, there is 
a final interval  $\delta_1(Q)\le
\delta<\delta_0(Q)$ where
bands with opposite Krein signature overlap, and as a consequence 
{\em all commensurate  SWs are  
unstable for small but nonzero amplitude}. 
This can be proven \cite{Anna} by considering the linear limit 
($\delta=\delta_{0}(Q), \psi_{n}^{2}=0$), where the spectrum of 
(\ref{eigeneqharm2}) is given by  $\omega_{l}=\pm 2 C
|\nu_Q(q)|$ with $\nu_Q(q)=\cos Q -\cos q$. In this limit, 
eigenvalues with opposite Krein signature 
(given by $\mbox{sign}(\nu_Q(q))$ \cite{Anna}) will overlap in  an 
interval around $\omega_l = 0$. Since for commensurate SWs  
a finite number of bands with nonzero widths appear for
$\delta<\delta_{0}(Q)$ when  $\psi_{n}^{2} \neq 0$, a 
continuity argument \cite{Anna} yields that bands with 
opposite Krein signature will continue to overlap close to $\omega_l = 0$ and 
cause instabilities until $\delta$ reaches 
some value $\delta_{1}(Q) < \delta_{0}(Q)$. (In the linear limit, the 
SW is stable in spite of the band overlap, since the eigenmodes are 
uncoupled.)

By contrast, for small-amplitude (analytic) incommensurate SWs the 
Cantor-like nature of the spectrum  of 
(\ref{eigeneqharm2})  (with 'bands' of zero width) prevents the use of a similar argument for  proving 
 instability. However, the analysis 
outlined  below shows \cite{Anna} that also these SWs are
unstable,  but the instabilities may be of higher order and thus more 
difficult to detect numerically or experimentally. 

The instability of analytic SWs is proven using the method of 
'band analysis' \cite{Aub97}. 
Its basic idea is to embed the eigenvalue problem (\ref{eigeneqharm2})  for
a non-Hermitian operator 
into a wider eigenvalue problem for a
Hermitian operator, which we choose as  (with  $\omega$  a real 
parameter)

\begin{equation}
\label{matrix}
\left ( \begin{array}{c} {\cal L}_1 \;\; \; -\omega \\ -\omega
\;\; \;
{\cal L}_0  \end{array} \right) \left ( \begin{array}{c} \{u_n\}\\\{v_n\}
\end{array}
\right)  = E \left ( \begin{array}{c} \{u_n\}\\\{v_n\} \end{array}
\right) .
\end{equation}
Then, each  eigenvalue $E_{\nu}$ is a smooth, real 
function ('band') of $\omega$. The real eigenfrequencies
$\omega_{l}$ of (\ref{eigeneqharm2}) are given by the intersections
of these bands with $E=0$, and an instability occurs whenever a band loses a 
pair of intersections with 
$E=0$, i.e., when a 'gap' opens around $E=0$. 

In the linear limit ($\psi_n \rightarrow 0$, $\delta \rightarrow
\delta_0(Q)$) the bands  can be labelled by a wave number $q$ and describe 
 straight lines 
$E_\pm (q; \omega) =  2C (\cos Q - \cos q) \pm \omega$. 
Thus, for each $q$ there are two bands  
(doubly degenerate since $E_\pm (q; \omega) = E_\pm (-q; \omega)$), and 
bands with different $q$ intersect when
$E_- (q_-; \omega)= E_+ (q_+; \omega)$.
When $\psi_n$ describes an analytic SW with small amplitude $A$, 
$\psi_{n}^2$ can be treated as a perturbation and expanded 
as $\psi_{n}^{2} = \sum_{p}f_{p} e^{i p(2n Q +\phi)}$, 
where the coefficients $f_p$ are of order $A^{2|p|}$ for $|p| \geq 1$, and 
thus exponentially decaying for $p\rightarrow \infty$. 
This introduces a coupling between eigenvectors 
of (\ref{matrix}) with wave numbers  $q$ and  $q+2pQ$ with coupling 
strength proportional to $f_p$, and 
gaps of width $\Delta E \propto 2|f_p|$ will open at each intersection
where $E_{-}(q;\omega)= E_{+}(q+2pQ;\omega)$. If these gaps surround 
the axis $E=0$, instabilities occur for the SW. 

\begin{figure}[htbp]
 \noindent
 \includegraphics[height=7cm,angle=270]{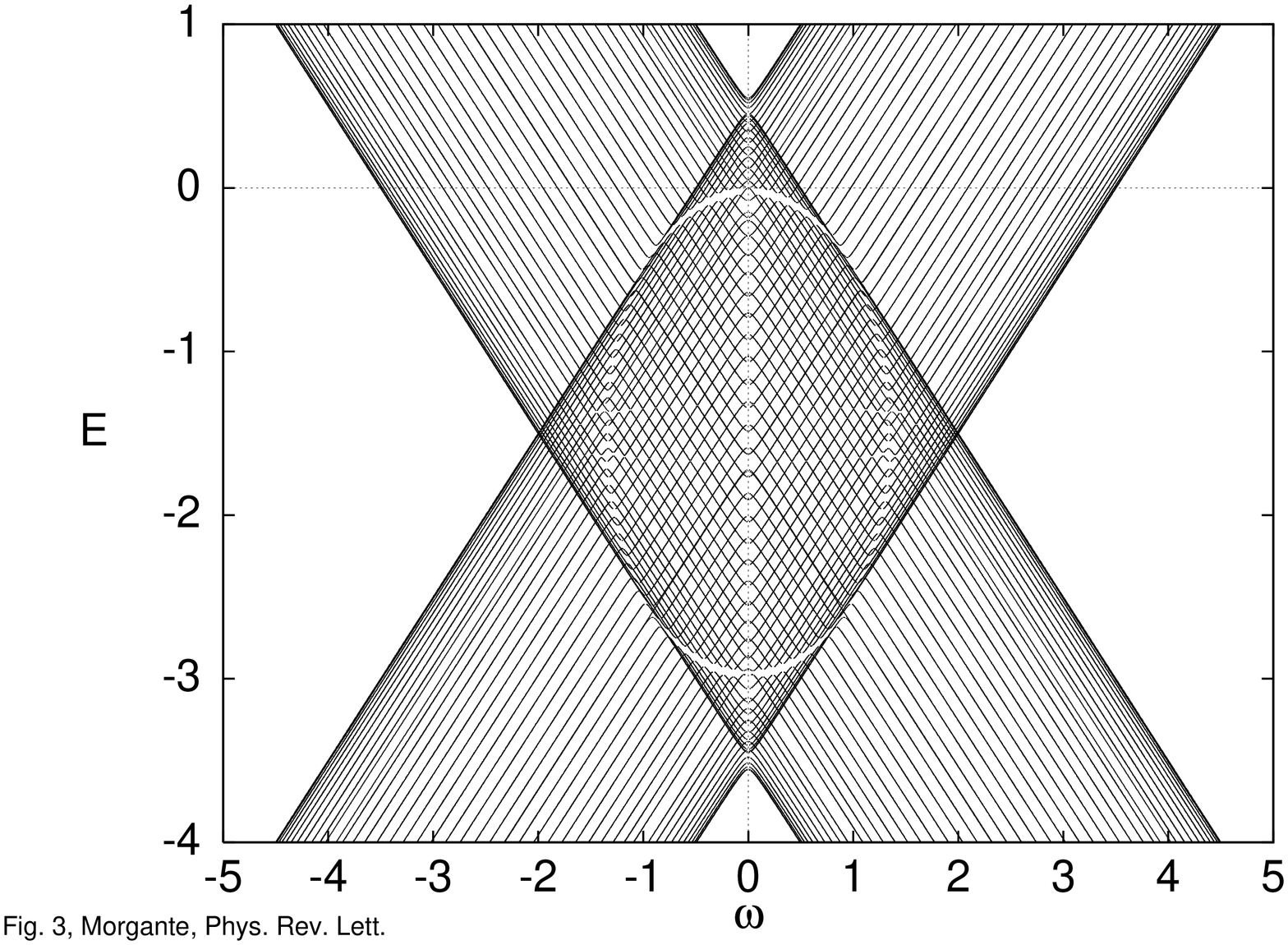}
 \caption{Band spectrum of (\ref{matrix}) for a SW 
 with $Q/2\pi = 34/89  \simeq \sigma_G $ ($Q > \pi/2$) and
 $\delta^\prime=3.4$ ($C=1$). The visible gap
 openings occur along the ellipse (\ref{ellips}) with  $p=1$.}
 \label{crossings}
 \end{figure}

\begin{figure}[htbp]
\noindent
\includegraphics[height=7cm,angle=270]{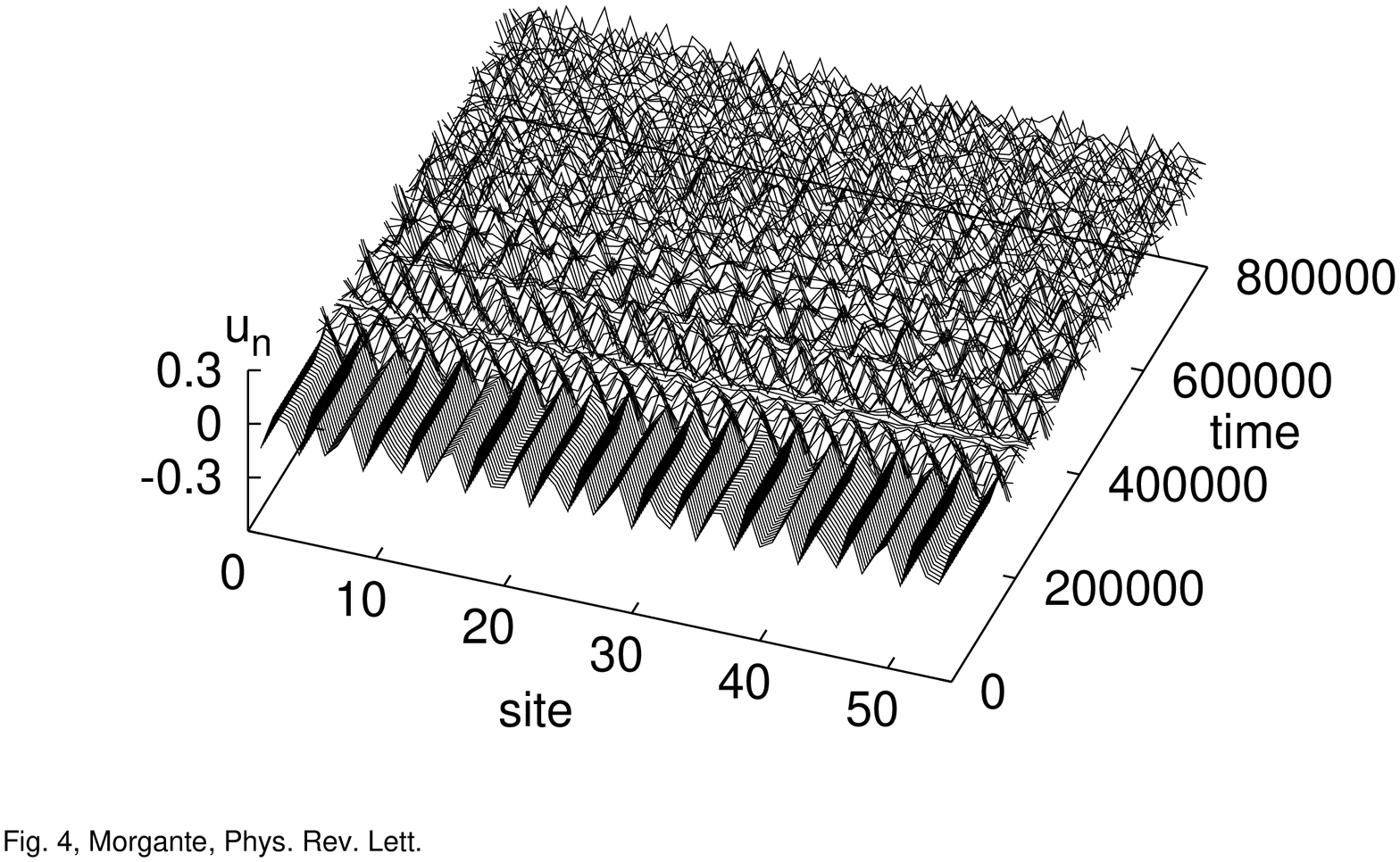}
\caption{Time evolution of an analytic SW (at integer multiples of the 
original period),
 perturbed only by numerical truncation 
errors, 
in a KG chain (\ref{DKG}) with $C_K=0.03$, $\omega_{b}=1.044$, 
and other parameters as in Fig.\ \ref{transition}. }
\label{motion}
\end{figure}

For fixed $p$ these
intersections describe  an ellipse in the $(\omega,E)$-plane given by
 \begin{equation}
 \left(\frac{\omega_p}{2 C \sin pQ}\right)^2+\left(\frac{E_p-2C\cos Q}
{2C \cos pQ}\right)^2 = 1 .
 \label{ellips}
 \end{equation}
Numerical calculations of the set of bands $E(\omega)$ confirm the existence
of these ellipses of gaps (see Fig.\ \ref{crossings}).
The ellipse with $p=1$ corresponds to the largest gaps (and thus the
strongest instabilities if the gaps surround $E=0$) and is for 
$\delta=\delta_0(Q)$ tangent to $E=0$ 
at $\omega=0$, lying below $E=0$ for $\pi/2<Q<\pi$ 
and above for $0<Q<\pi/2$.
Considering also the shift of the bands caused by the static term 
 $f_0$, 
we find \cite{Anna} that when
 $0<Q<\pi/2$  gaps will open around $E=0$ in the lower part of the
ellipse, and thus  {\em these waves become immediately unstable to
first order} (as for commensurate SWs with  $Q \neq \pi$ \cite{Anna}).
\vspace{-0.5cm}
By contrast, for irrational $Q$ with $\pi/2<Q<\pi$ all first-order gaps 
open {\em strictly below} $E=0$ (Fig.\ \ref{crossings}), so that 
{\em no first order instabilities} develop for these SWs. However, as is
seen from (\ref{ellips}), the ellipse corresponding to the $p$th order
intersections will intersect $E=0$ if $|\cos pQ|$ $>$ $|\cos Q| $, and
therefore {\em there is always a $p$ such that the SW is unstable to
order
$p$ for arbitrarily small amplitude}. But as the width of the gaps 
and the maximum instability growth rates are proportional to $|f_p|$,
the instabilities become very weak 
when the smallest $p$ yielding an instability 
becomes large (for $Q$ close to $\pi$). We
might therefore view these analytic SWs as 'quasi-stable'.
A typical example of the dynamics resulting from this instability in a KG 
chain is shown in Fig.~\ref{motion}, where, after the initial 
oscillatory dynamics and an intermediate regime of mainly translational 
motion, an apparently chaotic final state appears.  
Further details concerning
the long-time dynamics of the unstable SWs will be published
elsewhere.
\vspace{-1.0cm}
In conclusion, we have shown that for a general class of anharmonic 
lattices, different types of SWs with spatial period commensurate or 
incommensurate with 
the lattice exist,  but are generically unstable through oscillatory 
instabilities for small amplitudes. Thus, in discrete systems, the 
nonlinearity-induced coupling between counter-propagating waves 
leads to wave  break-down for arbitrarily weak anharmonicity {\em even if 
the individual propagating waves are modulationally stable}. The general 
nature of this result suggests that these instabilities should be observable 
in macroscopic as well as microscopic contexts. In particular, as was 
recently experimentally verified \cite{waveguides}, nonlinear 
optical waveguide arrays provide a direct application of the DNLS equation, 
and thus these systems are also good candidates for detection  of SW 
instabilities. 

M.J. acknowledges a EC Marie Curie fellowship.

\end{multicols}

\end{document}